# RISTRETTO: Assembly and Testing of the Seven-spaxel, High-resolution, Diffraction-limited Spectrograph

Leonie Hoerner[a], Bruno Chazelas[a*], Nathanaël Restori[a], Christophe Lovis[a], Nicolas Blind[a], Ludovic Genolet[a], Ian Hughes[a], Michael Sordet[a], Robin Schnell[a], Anthony Carvalho[a], Maddalena Bugatti[a], Adrien Crausaz[a], Samuel Rihs[a], Gilles Simond[a], Isaac Bernardino Dinis[a], David Ehrenreich[a], Emeline Bolmont[a], Christoph Mordasini[b], Martin Turbet[c], Angelie Alagao[d], Vincent Chambouleyron[d], Mathieu Motte[d], Arnaud Striffling[d], Francissco Oyarzun[d], Benoit Neichel[d], Jean-François Sauvage[d,e], Thierry Fusco[d,e].

[a]Observatoire de Genève, University of Geneva, 51 chemin de Pegasi 1290 Versoix, Switzerland,
[b]Physikalisches Institut, Universität Bern, Gesellschaftsstrasse 6, CH-3012 Bern, Switzerland,
[c]Laboratoire de Météorologie Dynamique, IPSL, CNRS, Sorbonne Universitée, 4 place Jussieu, F-75252 Paris Cedex 05, France,
[d]Laboratoire d'Astrophysique de Marseille (LAM), UMR 7326, Aix-Marseille Université, CNRS, CNES, 38 rue Frédéric Joliot-Curie, 13388 Marseille Cedex 13, France,
[e]ONERA - The French Aerospace Lab, 92322 Châtillon, France

## ABSTRACT

The RISTRETTO[1,2] project aims at the direct detection of the reflected light of extra-solar planets to measure albedos and detect possible biosignatures, using the high-contrast / high-resolution method. We report on the assembly, lab-testing and on-sky testing of the seven-spaxel high-resolution single-mode spectrograph which was built ahead of the rest of the instrument. The spectrograph is a high resolution echelle spectrograph build for high spectral fidelity being uder vacuum and thermally controlled. Once the assembly has been completed we had the chance to test it on sky using OHP 1.52 m telescope using the PAPYRUS AO system for injection in our spectrograph.

## 1. INTRODUCTION

RISTRETTO[1,2] is a project to build an instrument tailored at the detection of the reflected light of exoplanets, and in particular to detect, in the visible spectrum, the reflected light of Proxima b, the closest terrestrial exoplanet from earth. In order to achieve such a feat, one needs to overcome two major obstacles. The first is *angular resolution*, in order to resolve the planet from the star one need an adaptive optic system on an 8m class telescope. The second is *contrast*, in order to detect Proxima b (if it was an Earth-like planet) the contrast would be of $10^7$ between the star and the planet. To address this, the instrument uses a coronagraph as a first stage to gain a factor $10^4$, and in cascade a high resolution, high spectral fidelity spectrograph in order to gain the rest of the contrast. This last element uses the radial velocity difference between the planet and the star to disentangle the spectrum of the light. The development of the project has had 2 phases. As the construction of the required spectrograph[3] was easier, we started with it, while doing R&D to develop the necessary technology to build the AO[4–6] and coronagraphic[7] system. The construction[8] of the spectrograph was concluded in January 2026, and it immediately left for some test on the 1.52 m telescope at OHP thanks to the PAPYRUS AO system.
The spectrograph is a seven-spaxel diffraction limited spectrograph. Its resolution is 135'000. Its design is similar to HARPS: an echelle spectrograph under vacuum and thermally controlled.

* bruno.chazelas@unige.ch

## 2. OPTO-MECHANICAL ASSEMBLY AND INTEGRATION OF THE SPECTROGRAPH

The spectrograph opto-mechanics has been designed in aluminum. All parts have been designed to hold optics in a way that maintains the optical quality of the optics, keeps the instrument as stable as possible when thermally controlled, and withstands the transport while keeping everything assembled. We made extensive use of canoe-sphere as a means to provide reproducible alignment with positioning repeatability in the 10-15 micron range and the angular reproducibility in the arcminute range. When degrees of freedom were necessary for the adjustment of the optical element, the adjustments were performed with shimming, with two exceptions: the main echelle grating and the cross disperser were mounted using a scheme inspired by a design from ESO[9].

The alignment was performed using an alignment telescope for the main part. Shim thicknesses were determined iteratively using computer software. A particular step in the alignment has been the alignment of the cross-disperser relative to our main alignment reference target. This has been performed out of the spectrograph optical bench, on a specific alignment fixture. To meet our requirement for an alignment precision of ~1 arcminute, we used a FARO arm to reproduce the orientation of the alignment reference mirror on the cross-disperser mount. The 3d measurement were fed to a software that computed the offset to apply to 6 small actuators in order to bring the prism to its final orientation. With an iterative process, we could reach the desired position.

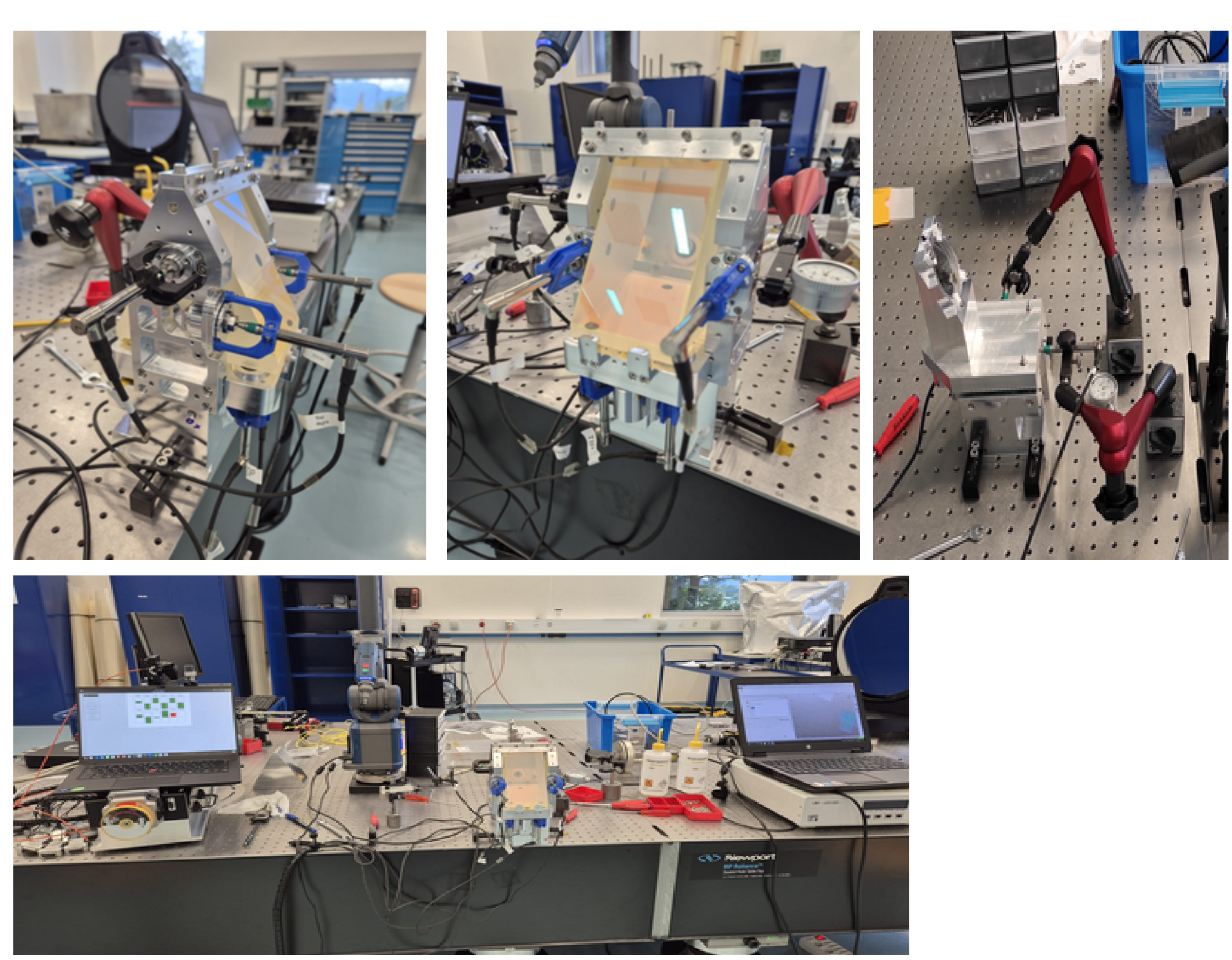

Figure 1: Photo of the alignment setup for the Cross disperser prism. The prism is co aligned to an alignment target on the same isostatic interface. A Faro arm, 6 digital dial indicators have been used. The precision achieved is ~1 arcmin

The last part of the alignment has been done optically using the detector. It was necessary to adjust the fiber bundle orientation in order to adjust the spacing between the traces of the different fibers. The blaze has been centered and the spectral orders have been centered directly using the images of the detector.

The last part of the alignment was the focus. It has been first set at ambient pressure, then offset to account for the vacuum based on the analytical model. Lack of time before the departure for the on-sky tests did not allow for the finest adjustments; in particular the tilt of the detector has not been adjusted. The characterization of the PSF/resolution is shown in the Performance section of this paper. We are still analyzing the performances of the instrument to figure out what is the best course of action. A refocus and tilt of the detector is very probable.

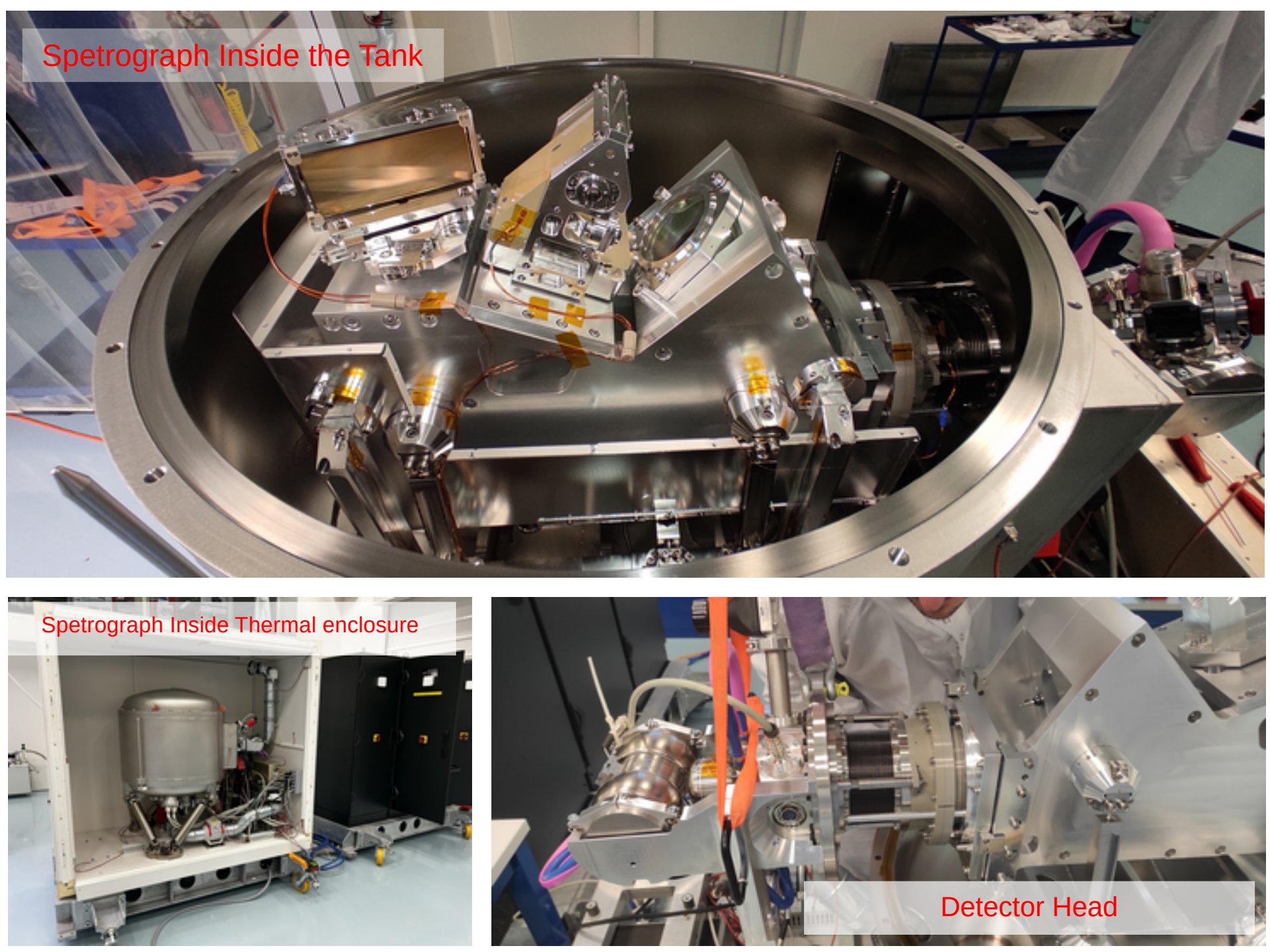

Figure 2: Photos of the Spectrograph internals. To view the top of the spectrograph where one can see the Echelle Grating the Cross disperser Prism and the Main collimator, on the bottom left the thermal enclosure just before being closed and on the bottom right the Detector head featuring the flexible bellow that allow the detector to be rigidly fixed to the optical table but flexible with respect to the vacuum tank.

## 3. FIBER INJECTION MODULE FOR OHP

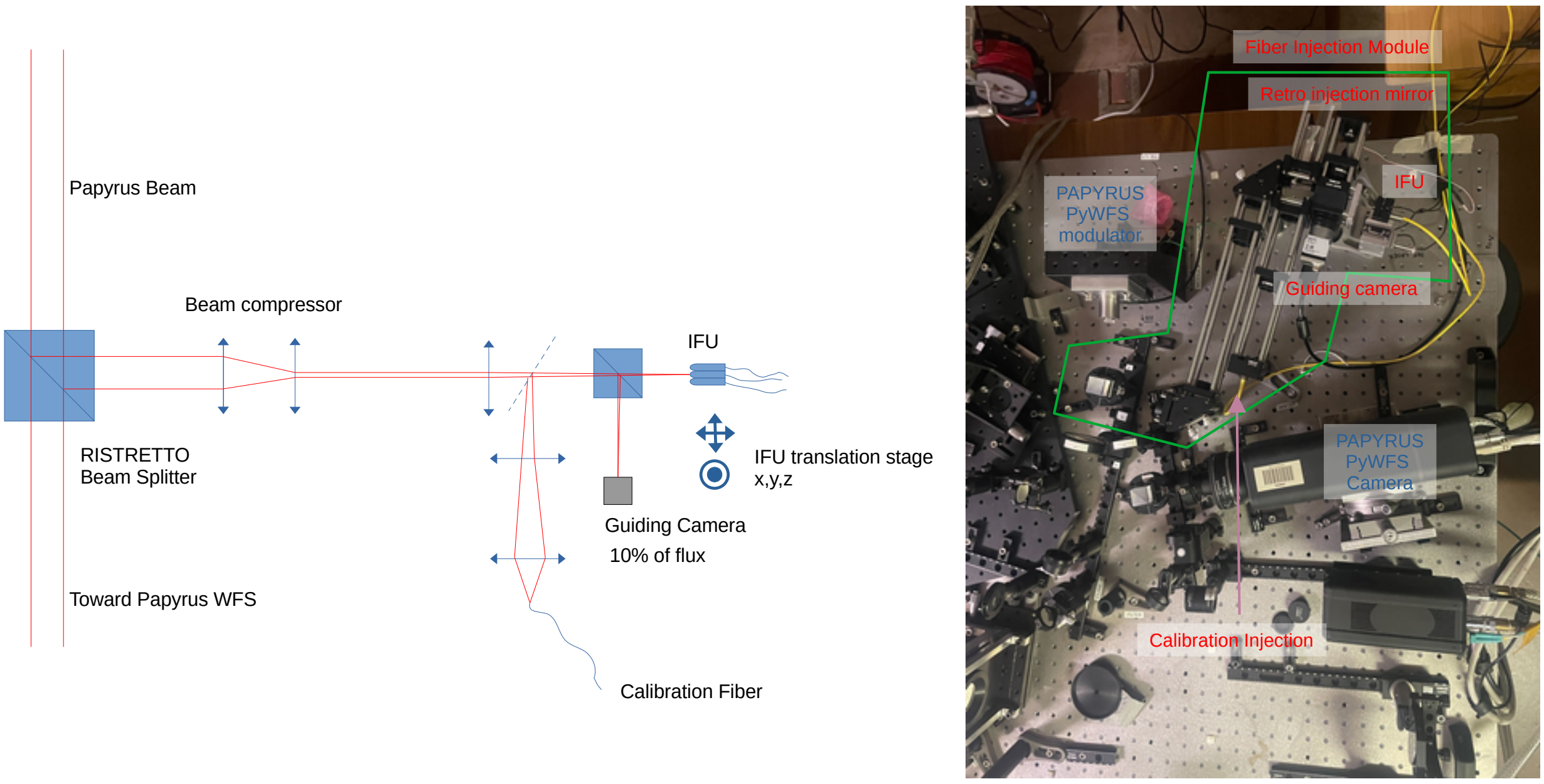

Figure 3: Scheme and photo of the Fiber injection module used at OHP for the on sky tests.

In order to test the spectrograph on-sky we used the PAPYRUS experiment at OHP. It has been possible to insert a fiber injection module for the spectrograph in just before the WFS modulator mirror. Figure 3 shows the optical scheme of the fiber injection module.

The functions of this module are:

- Receiving the light of the PAPYRUS AO,
- A guiding camera ensure the centering and guiding the light on the IFU,
- Inject the spectrograph calibration in the IFU,
- A retro-injection mirror allows finding the position of the IFU on the guiding camera when light is retro-injected on the fiber bundle.

## 4. CONTROL AND SOFTWARE

The control of the experiment is mostly done with a Beckhoff PLC, reading and controlling all the sensors and implementing the thermal control loops. A Linux workstation deals with the acquisition of the detector frames producing the FITS images gathering all the necessary metadata, reading the exposuremeter and adding the time series to the FITS image. The architecture of the software is shown on the Figure 4. We use 2 off the shelf databases, one for time series (Influx) and one for real-time values (Redis). Most of the pieces of the software run in containers are written in Python and are interconnected via REST APIs. We use Grafana as a tool to monitor the experiment, having different dashboards, for the different parts of the experiment.
The fiber injection module software is running on a second machine.

The data reduction pipeline for the spectrograph is under construction the result shown in the performance section are still preliminary.

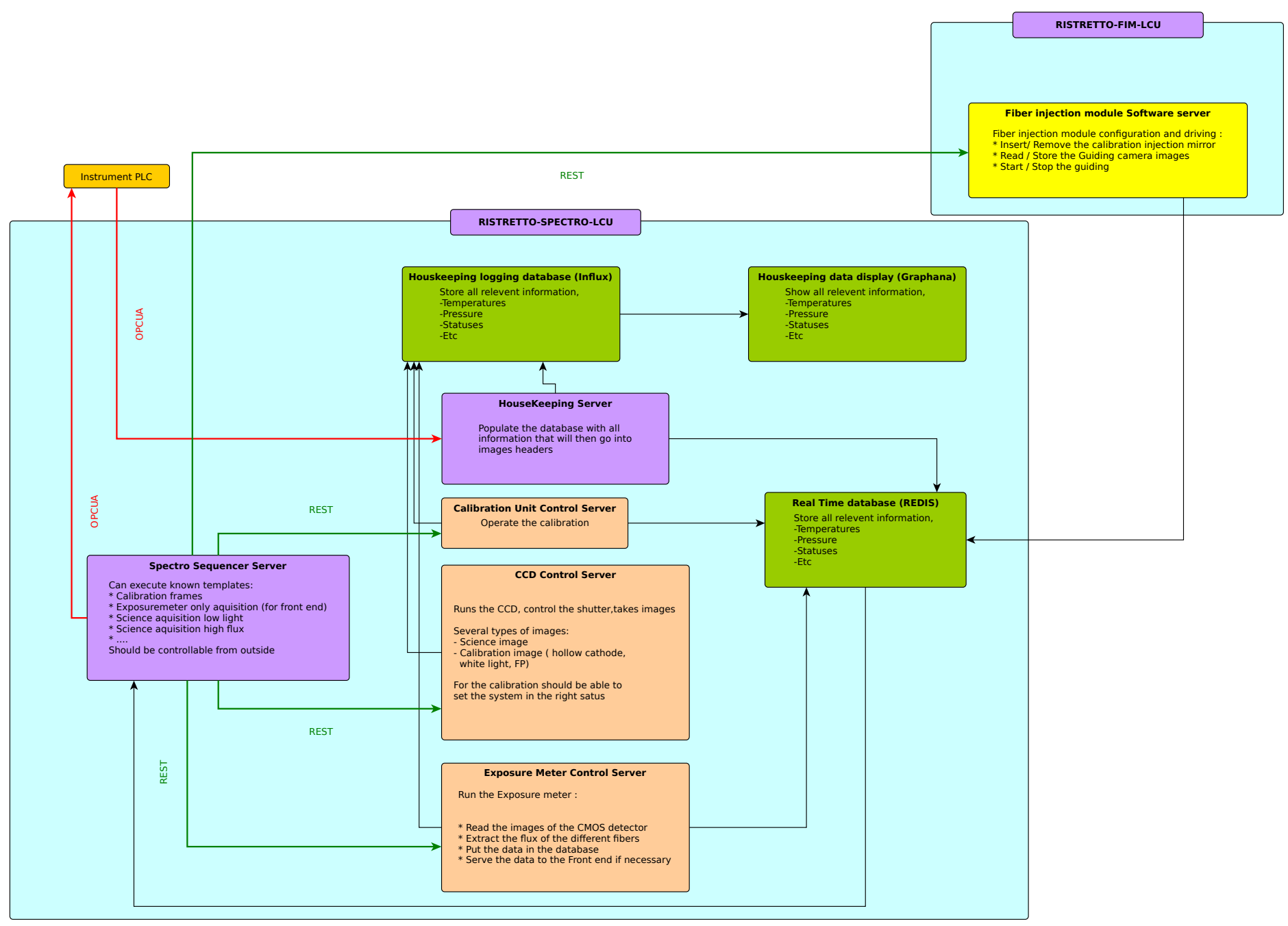

Figure 4: Scheme of the RISTRETTO spectrograph Software

## 5. TRANSPORT AND INSTALLATION AT OHP

The whole instrument has been installed in a marine container (Figure 5) that has been insulated, and air-conditioned. The spectrograph is installed fixed on the container floor with wire-rope insulators. It allowed a safe transport to and back from OHP with the fully assembled/aligned instrument. The control cabinets were also well fixed to the ground of the container. At OHP the container was kept as is and connected to the electricity, while the fiber link was passed inside the telescope up to the FIM hosted by the PAPYRUS experiment. As a consequence of this choice, the first spectra have been taken only one day after the arrival to OHP. It took however additional time to reach operational thermal stability. We could spot bringing the spectrograph back from OHP to Geneva a small misalignment of ~20 pixels in the dispersion direction and a small defocus. This will be studied and fixed in the next test phase for the spectrograph.

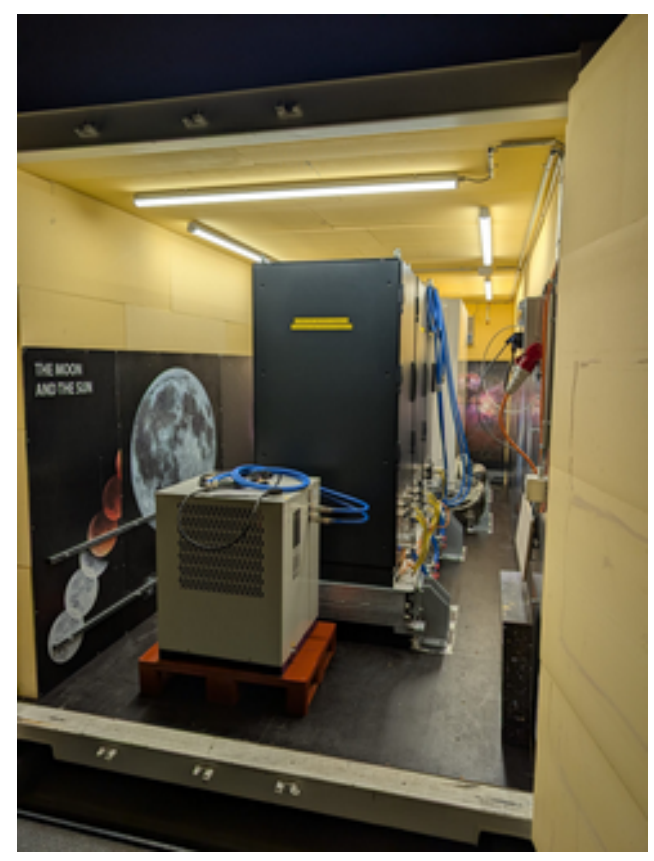

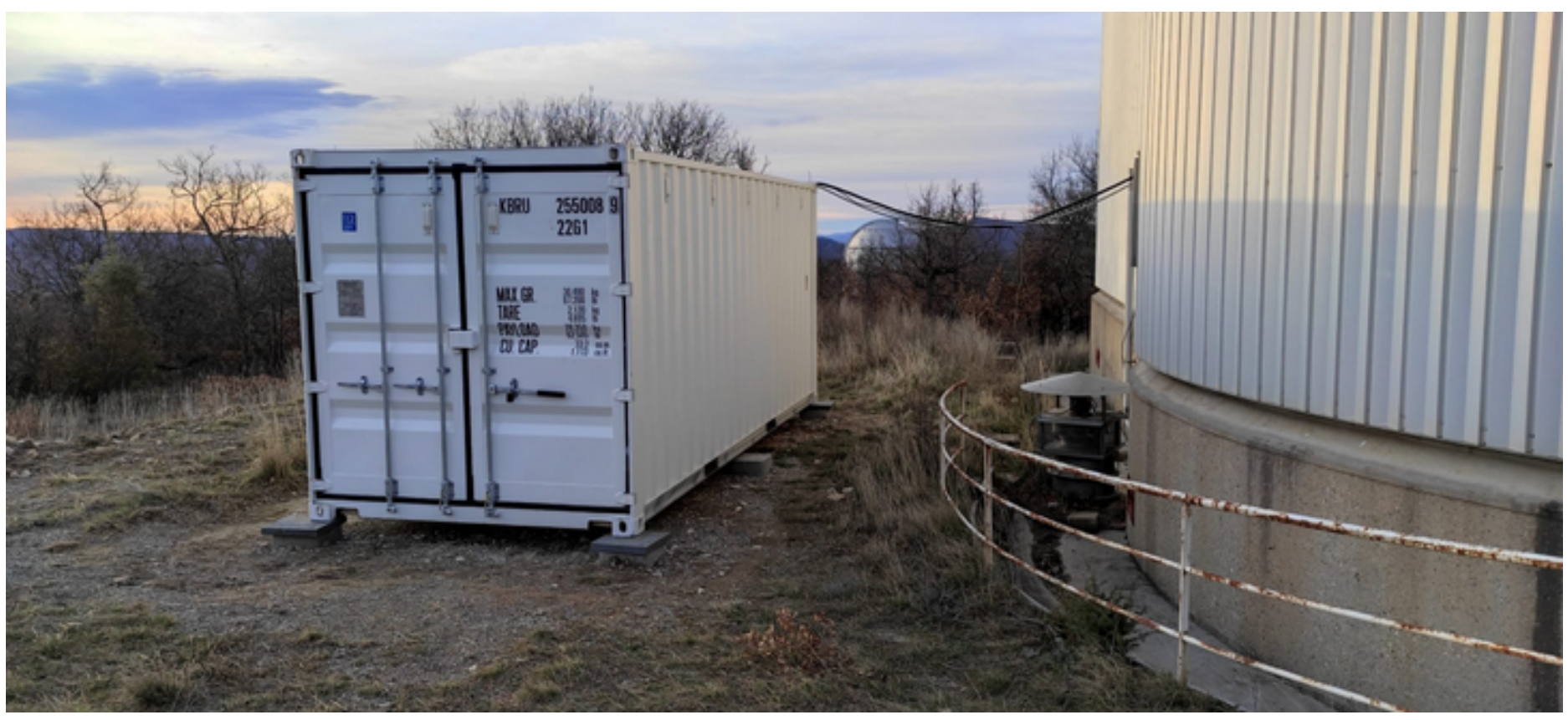

Figure 5: The RISTRETTO spectrograph in its container, Installed on the 1.52m at OHP

# 6. PERFORMANCE

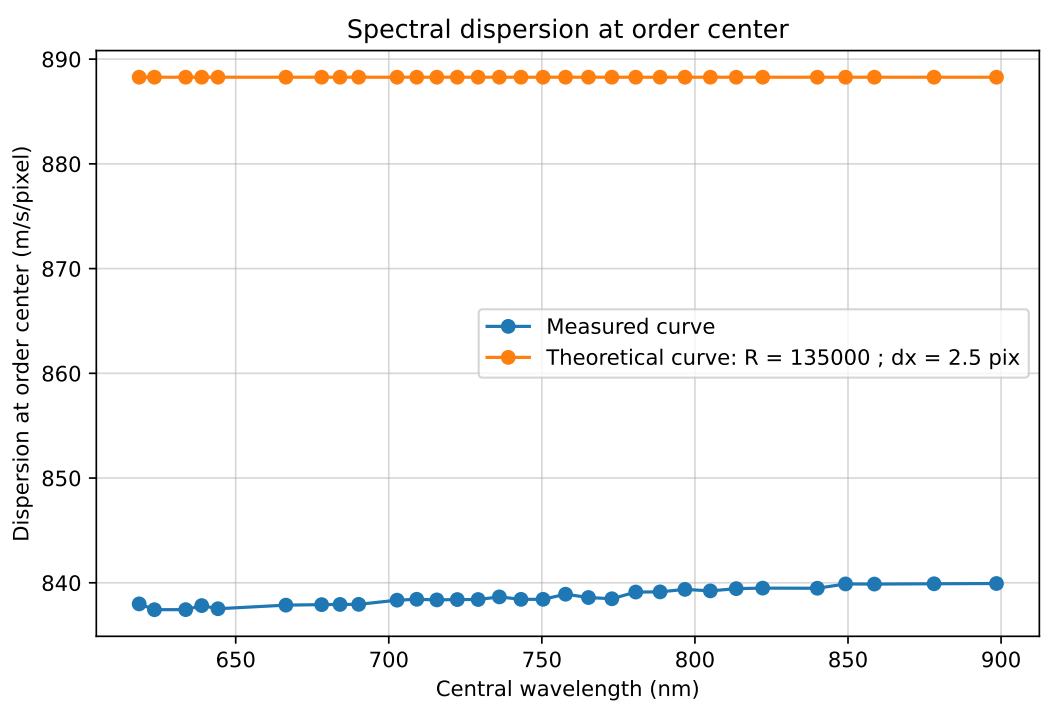

Figure 6: Dispersion at the center of orders: size of a pixel in m/s.

**Efficiency**

- The wavelength range covered goes from 617 nm to 860 nm, where there is an important efficiency drop due to the lack of guiding by the optical fibers.
- The efficiency for the best fiber of the spectrograph made in the lab at 635 nm, for the best fiber is >37%±7%. (83% ±5% for from fiber input to after FM0, 49% ±5% from FM0 to before detector, 91% for the detector efficiency). The loss in the fiber train is probably overestimated as was tested with direct FC/PC connections, We suppose that the measure is done at the maximum of the blaze because it was performed at a moment where the orientation of the grating was not well-defined. The measurement is in the proper range we consider it a lower bound of the actual performances.
- The efficiency of a similar the prototype IFU as the one used at OHP is ~62% on the central fiber (and between 54 and 60 % on the other fibers. We have not yet had the occasion to confirm the values with the one actually used at OHP.
- The injection of light with the PAPYRUS had some additional losses due to the lack of ADC. See Figure 7.

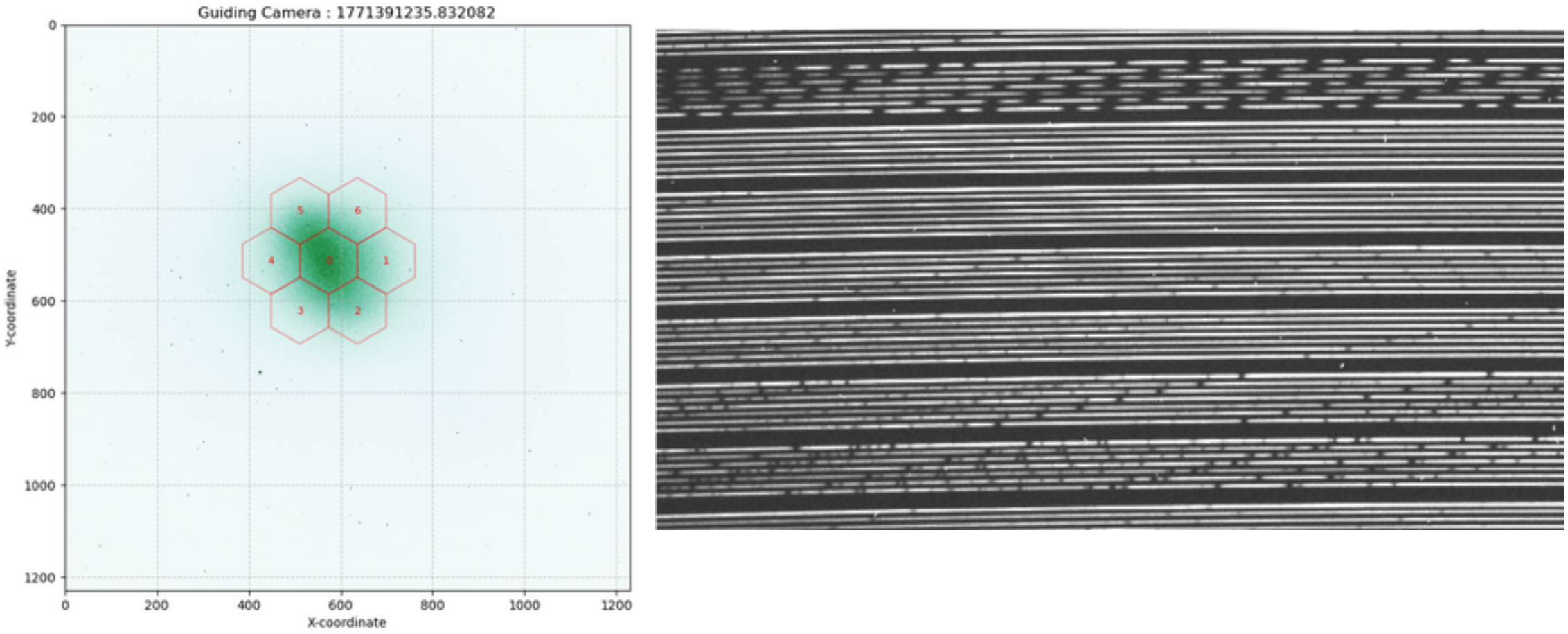

Figure 7: On the left, a star as seen on the FIM guiding camera, with the IFU spaxels represented as an overlay. The center of the outer spaxels are located at 2λ/d. The elongation of the PSF is due to the lack of ADC. It is a 10s exposure. On the right an extract of the spectral format, showing Arcturus in the region of the oxygen band. One can clearly see the 7 fibers and how they are arranged in a nearly horizontal slit in order to set the required distance between the different fibers.

- The efficiency of the injection into the IFU has not been fully established. We can however estimate it combining the images from the guiding camera and the exposure meter. A first look at the data show at least 10% of the light has been injected in the spectrograph (combining all the spaxels) on one hour exposure. This coherent with a fiber link efficiency of 83% and IFU efficiency of ~ 60%m and a strehl ratio of ~20%.

**PSF size**

PSF size (Figure 8) are varying across the detector. Currently, given the lack of time to align better the detector there is a large variation in the PSF size. It is overall too small and under sampled. As the spectrograph is back in Geneva this will be further studied and improved, There are 2 degrees of freedom to improve the situation: focus and tilt.

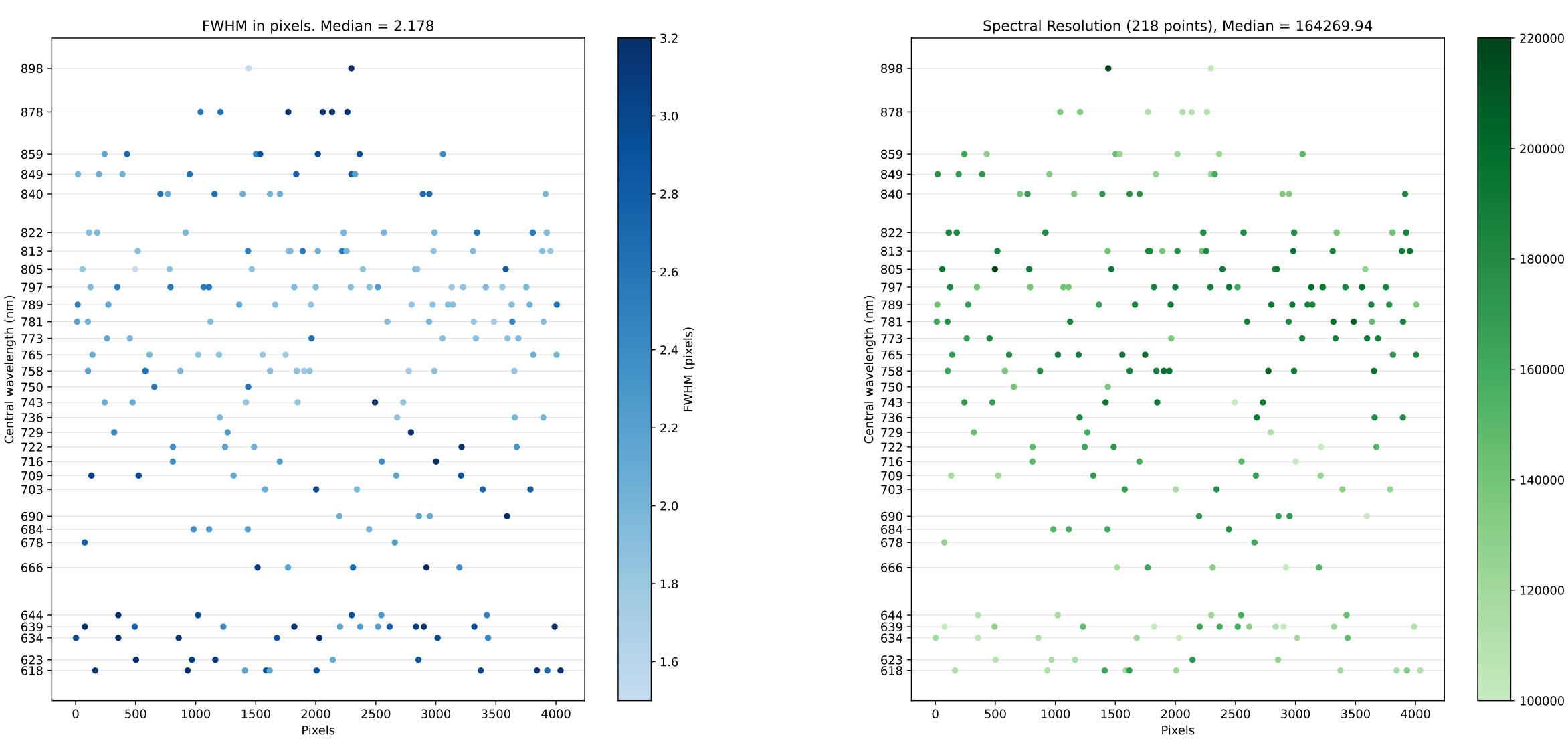

Figure 8: Map of the extracted PSF size on the detector, using UNe exposures, and map of the spectral resolution.

**Dispersion**

The full wavelength solution is not yet completely established. But preliminary study we are in line with expectations at the center of the order, see figure 6. There is a 5.6% difference between the experimental curve and the design value, it is easily explained by the different tolerance on the focal length of the camera and the blaze angle (and thus the actual incidence angle).

**Stability**

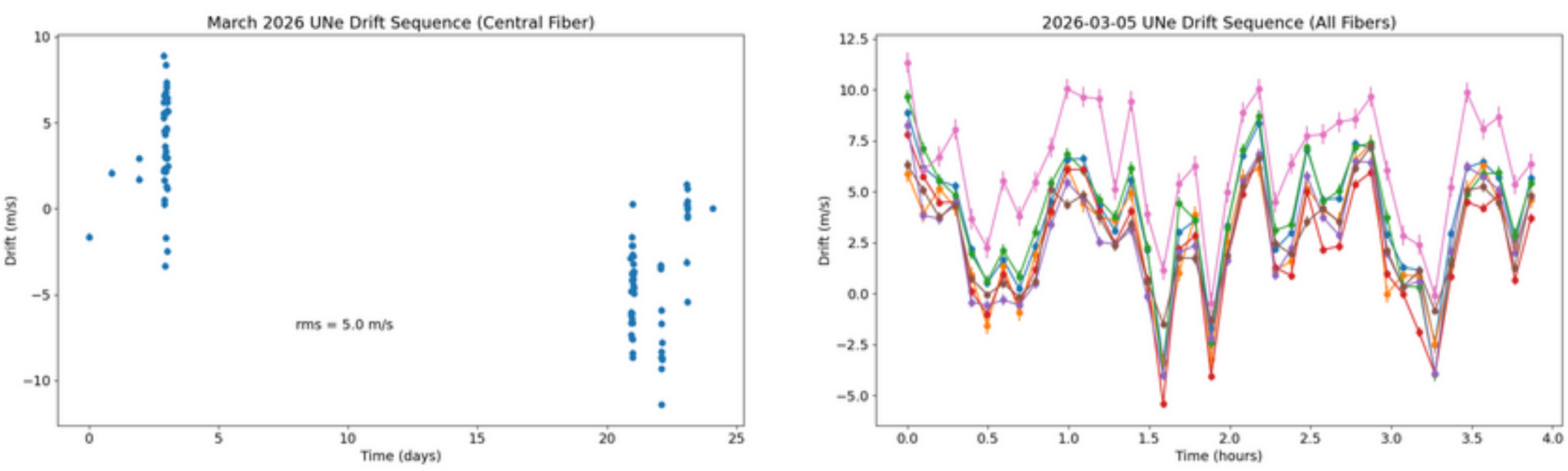

Figure 9: Absolute Stability of RISTRETTO spectrograph, during its stay at OHP. The measurement is done with UNe exposures of 300s. One see instabilities at the level of 3 m/s on short time scales. One also sees that the 7 fibers see very similar effects within 1.1 m/s rms over 4 hours.

- The thermal stability of the spectrograph was in line with expectations or even better as we were in a more stable environment than the one for which the instrument is built for (the container that hold the instrument is air-conditioned). Over the month of March 2026 the optics temperature (Grating/Input fiber / Cross Disperser) was better than 1.4 mK
- The thermal stability of the detector in this phase was not great as there is a noise picked by the thermal sensor of the detector when the thermal control loop of the spectrograph is on. When off we could reach ~1 mK rms stability in idle mode while the stability reached once the instrument stabilization system was online was ~50 mK rms.
- The pressure in the instrument has been maintained at below $10^{-4}$mBar for one month using continuous turbo pumping. The pressure actually decreased down to 6 $10^{-5}$mBar over one month. This however would have a negligible effect on the instrument stability.
- The measured absolute stability of the spectrograph using UNe spectra show several things:
    - There is a long time trends over a month of measurement of ~5 m/s.
    - Short time absolute stability over 4 hours is 2.8 m/s.

These values are well within our requirement of 10 m/s. The instabilities arise on very short time frame (few minutes) and extremely correlated on the 7 fibers of the spectrograph (as shown on the Figure 9)

Given this we think that the overall stability should improve significantly once we have tracked the short terms instabilities (possible by correcting the detector thermal stability issues)

**Ghosts**

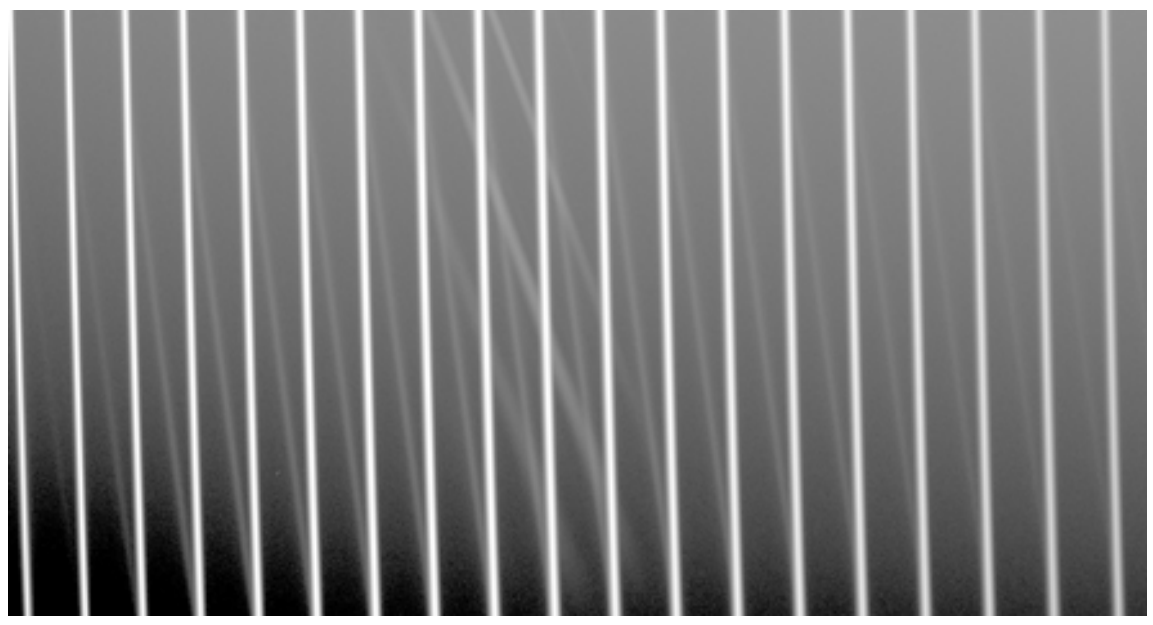

Figure 10: Ghost orders visible in with white light. There are in focus ghost orders that come from spurious reflection in the field lens of the detector. Here only one fiber is illuminated and the diffraction orders are displayed vertically.

In white light we can see clearly 2 series of in focus ghost orders. They are reflection on the Field lens and the detector. Their intensity is lower than 0.5%.

## 7. CONCLUSION

This paper presents the current the first on sky test of the RISTRETTO spectrograph. The first results are very encouraging, The spectrograph subsystem is now mostly complete. We already see that the most important parameters are in specifications. Until the end of 2026 we plan to :

- Do further tests,
- Improve the Data reduction software,
- Improved optical alignment (Focus),
- Complete automatization,
- Install a single-mode Fabry-Perot for calibration,

## 8. ACKNOWLEDGMENTS

This work has been carried out within the framework of the National Centre of Competence in Research PlanetS supported by the Swiss National Science Foundation under grants 51NF40_182901 and 51NF40_205606. The RISTRETTO project was partially funded through the SNSF FLARE programme for large infrastructures under grants 20FL21_173604 and 20FL20_186177. The authors acknowledge the financial support of the SNSF.

The RISTRETTO project acknowledges the financial support of the Swatch Group.

The RISTRETTO team is deeply grateful with the OHP team, Marc Ferrari, François Dolon, Françcois Huppert and Gregory PINA. the same extends to the PAPYRUS team that was of enormous help for our tests.